\documentclass[sigconf,natbib=true]{acmart}

\AtBeginDocument{%
 \providecommand\BibTeX{{%
 \normalfont B\kern-0.5em{\scshape i\kern-0.25em b}\kern-0.8em\TeX}}}

\setcopyright{acmcopyright}
\copyrightyear{2018}
\acmYear{2018}
\acmDOI{XXXXXXX.XXXXXXX}

\acmConference[Conference acronym 'XX]{Make sure to enter the correct
 conference title from your rights confirmation emai}{June 03--05,
 2018}{Woodstock, NY}
\acmPrice{15.00}
\acmISBN{978-1-4503-XXXX-X/18/06}

\newcommand{\ie}{\emph{i.e., }}
\newcommand{\eg}{\emph{e.g., }}

\newcommand{\wrt}{\emph{w.r.t. }}

\usepackage{enumitem}
\usepackage{verbatim}
\usepackage[ruled]{algorithm2e}
\usepackage{multirow}
\usepackage{subfigure} 
\usepackage{ragged2e}
\usepackage{caption}
\usepackage{graphicx}
\usepackage[normalem]{ulem}
\useunder{\uline}{\ul}{}
\usepackage{amsmath}
\usepackage{setspace}
\usepackage{comment}
\usepackage{bm}
\usepackage{makecell}
\usepackage{adjustbox} 
\usepackage{bbding}
\usepackage{color}

\title{Generative Multi-Target Cross-Domain Recommendation}

\author{Jinqiu Jin$^{1}$, Yang Zhang$^{2}$, Fuli Feng$^{1}$, Xiangnan He$^{1}$}

\affiliation{\institution{$^1$University of Science and Technology of China, $^2$National University of Singapore\country{}}}

\email{jjq20021015@mail.ustc.edu.cn, {zyang1580,fulifeng93,xiangnanhe}@gmail.com}

\begin{document}

\begin{abstract}
Recently, there has been a surge of interest in Multi-Target Cross-Domain Recommendation (MTCDR), which aims to enhance recommendation performance across multiple domains simultaneously. Existing MTCDR methods primarily rely on domain-shared entities (\eg users or items) to fuse and transfer cross-domain knowledge, which may be unavailable in non-overlapped recommendation scenarios. Some studies model user preferences and item features as domain-sharable semantic representations, which can be utilized to tackle the MTCDR task. Nevertheless, they often require extensive auxiliary data for pre-training. Developing more effective solutions for MTCDR remains an important area for further exploration.

Inspired by recent advancements in generative recommendation, this paper introduces GMC, a \textbf{G}enerative paradigm-based approach for \textbf{M}ulti-target \textbf{C}ross-domain recommendation. The core idea of GMC is to leverage semantically quantized discrete item identifiers as a medium for integrating multi-domain knowledge within a unified generative model. GMC first employs an item tokenizer to generate domain-shared semantic identifiers for each item, and then formulates item recommendation as a next-token generation task by training a domain-unified sequence-to-sequence model. To further leverage the domain information to enhance performance, we incorporate a domain-aware contrastive loss into the semantic identifier learning, and perform domain-specific fine-tuning on the unified recommender. Extensive experiments on five public datasets demonstrate the effectiveness of GMC compared to a range of baseline methods.
\end{abstract}



\maketitle

\section{Introduction}

\begin{figure}[t]
\centering
\includegraphics[width=0.85\columnwidth]{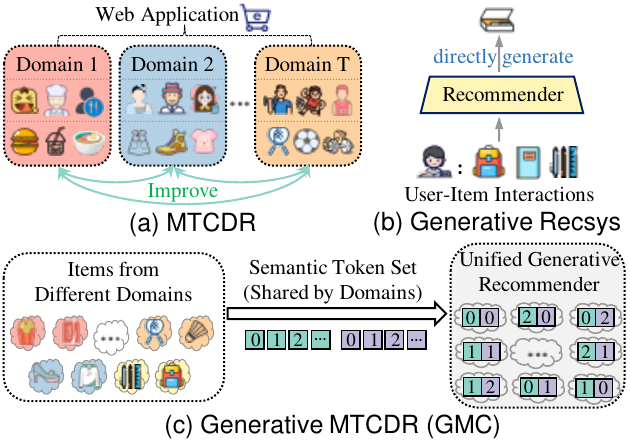}
\vspace{-8pt}
\caption{(a) Illustration of the MTCDR task. (b) Illustration of the generative recommendation paradigm. (c) The core idea of our proposed GMC approach.}
\label{fig:intro}
\vspace{-12pt}
\end{figure}

Recommender systems address the information overload issue by delivering personalized contents to users~\cite{zhang2019deep,wu2022survey}. In practice, most real-world applications consist of multiple sub-channels or sub-scenarios~\cite{zhang2024ninerec,huan2024exploring}, such as the product categories under the \textit{Shop by Department} section of Amazon. To this end, \textit{Multi-Target Cross-Domain Recommendation} (MTCDR) has gained increasing attention recently, which aims to improve recommendation performance across multiple domains simultaneously~\cite{zhu2021cross,zang2022survey} (Figure~\ref{fig:intro}(a)). 

A key factor for achieving promising MTCDR performance is effectively fusing and transferring recommendation knowledge, which relies on shared linkages across different domains. For example, existing methods leverage \textit{overlapped features} (\eg user, item or tag)~\cite{cui2020herograph,guo2023disentangled,xu2023neural,liu2024heterogeneous} across different domains as a bridge for the information integration and transfer. However, in practical applications, the item's content and active users in different scenarios may vary significantly, making it difficult to extract valid overlapped features between domains, and thus cross-domain recommendation becomes unfeasible. Recently, some methods leverage pre-trained language models to model user preferences and item features as semantic representations using \textit{textual descriptions}~\cite{hou2022towards,hou2023learning,li2023text,fu2024exploring}. However, they rely on auxiliary recommendation data for model pre-training to learn universal language representations, which is often unavailable. Without such pre-training data, the quality of learned representations may be suboptimal.

Recently, in the broader field of AI, significant advancements have predominantly adopted a generative approach~\cite{zhao2023survey,touvron2023llama,raffel2020exploring}, and recent recommendation works (outside the MTCDR field) are also transitioning to the generative paradigm in both academia and industry~\cite{wang2023generative,deldjoo2024review,zhai2024actions}. Unlike traditional \textit{discriminative} methods which calculate a matching score for each candidate item~\cite{kang2018self,wu2022survey,hou2023learning}, as shown in Figure~\ref{fig:intro}(b), \textit{generative} recommender systems directly generate the target item, simplifying the multi-stage recommendation process~\cite{li2024survey,liu2024recflow}. By reformulating recommendation as a context-aware generation task, recent works on generative recommender systems have demonstrated remarkable performance~\cite{geng2022recommendation,rajput2023recommender,zheng2024adapting,tan2024idgenrec}. These encouraging results raise the following question: \textit{How can the generative paradigm be applied to solve the MTCDR task?}

Building a generative recommender system generally involves two key aspects~\cite{li2024survey,li2024large,xu2024openp5}: \textbf{Item Identifier Learning}, which aims to uniquely and precisely represent each candidate item, and \textbf{Generative Model Training}, which focuses on learning the recommendation knowledge from the training data. To solve the MTCDR task in a generative manner, we propose that the following two requirements should be satisfied: (1) The identifiers of items in different domains should be sharable and linkable, enabling the integration of information from multiple domains. (2) Building on this, the generative model should be capable of fusing and transferring recommendation knowledge across domains. In addition, both the item identifier and generative recommender should ideally capture domain-specific characteristics of the data.

\begin{sloppypar}
Based on the above analysis, we propose a novel \textbf{G}enerative recommendation framework for \textbf{M}ulti-target \textbf{C}ross-Domain Recommendation, named \textbf{GMC}. As shown in Figure~\ref{fig:intro}(c), the core idea of our approach is to use discrete item codewords, derived from semantic quantization, as a bridge for cross-domain knowledge integration through training a unified generative recommendation model. In particular, GMC consists of three main steps: (1) For \textbf{Item Identifier Learning}, we use a unified Residual Quantization Variational Autoencoder (RQ-VAE)~\cite{lee2022autoregressive,zeghidour2021soundstream} to tokenize the semantic representations of items from different domains, thereby obtaining cross-domain shared hierarchical discrete semantic identifiers. Specifically, since items within the same domain are generally more similar to each other than items from different domains, we introduce a novel domain-aware contrastive loss that brings quantized embeddings of intra-domain items closer together. (2) For \textbf{Unified Recommender Training}, we formulate the next-item recommendation as a semantic token generation task~\cite{rajput2023recommender,singh2024better}, which is instantiated by training a unified sequence-to-sequence model using recommendation data from all domains. Given the differences in data characteristics across domains, we further perform (3) \textbf{Domain-Specific Fine-tuning}. Specifically, we incorporate domain-specific LoRA modules~\cite{hu2022lora} into the unified recommender and update the corresponding LoRA parameters using domain-specific data, while freezing all other parameters of the unified model. To validate the effectiveness of our method, we conduct experiments on five real-world datasets, where GMC outperforms several competitive baselines in terms of overall performance.
\end{sloppypar}

To summarize, the contributions of this work are three folds:
\begin{itemize}[leftmargin=*]
\item We thoroughly analyze the limitations of current MTCDR methods and, for the first time, discuss how the generative recommendation paradigm can be applied to address the MTCDR task.
\item We propose a novel method, GMC, whose key idea is to extract the semantic information of items from different domains into one set of discrete codewords, and then train a unified generative recommender to fuse and transfer cross-domain knowledge. 
\item We conduct extensive experiments on real-world datasets, demonstrating that GMC achieves superior multi-domain recommendation performance compared to several baseline methods.
\end{itemize}

\section{Related Work}

\noindent\textbf{Cross-Domain Recommendation. }
Cross-domain recommendation has emerged as an effective approach to address data sparsity and cold-start challenges by leveraging rich information from source domains to enhance performance in target domains~\cite{zang2022survey,zhu2022personalized,hu2018conet,li2020ddtcdr,xie2022contrastive}. A critical challenge lies in extracting and transferring the cross-domain information effectively. To tackle this issue, for instance, M2GNN~\cite{huai2023m2gnn} extracts content-based interests via tags and constructs a heterogeneous information network to model inter-domain relatedness; ALCDR~\cite{zhao2023beyond} employs an optimal transport based model to infer anchor links and aggregates domain-aware collaborative signals; AutoTransfer~\cite{gao2023autotransfer}, CUT~\cite{li2024aiming} and HJID~\cite{du2024identifiability} address the negative transfer problem by identifying the useful information in the source domain for transfer to the target domain. Additionally, MultiLoRA~\cite{song2024multilora} and MLoRA~\cite{yang2024mlora} incorporate domain-specific LoRA modules into the CTR models to effectively capture multi-domain recommendation knowledge.

In this work, we focus on a more general and practical problem known as \textbf{Multi-Target Cross-Domain Recommendation} (MTCDR), which aims to concurrently improve the recommendation performance across multiple domains~\cite{zhu2021cross,zhu2021unified,liu2024heterogeneous,li2024mutual}. Most existing MTCDR solutions leverage overlapping information between domains to integrate and transfer multi-domain knowledge. For example, HeroGRAPH~\cite{cui2020herograph}, DR-MTCDR~\cite{guo2023disentangled} and NMCDR~\cite{xu2023neural} utilize overlapping users as anchors and construct graph structures to model both domain-shared and domain-specific knowledge. AMID~\cite{xu2024rethinking} builds interest groups based on user behaviors and spreads information among both overlapping and non-overlapping users. RUR~\cite{chen2025recurrent} introduce user association representations as containers for domain features to transfer the information obtained through a recurrent-balance learning approach. However, these methods are inapplicable in more general scenarios where such domain-shared entities are unavailable. Other approaches such as UniSRec~\cite{hou2022towards}, VQ-Rec~\cite{hou2023learning} and RecFormer~\cite{li2023text} can be adopted to solve the MTCDR task, as they model user preferences and item features as language representations, which can serve as the bridge for cross-domain knowledge transferring. However, they rely on auxiliary domains for semantic representation pre-training, which requires large data size that may be unavailable. In contrast, we aim to use the generative paradigm for MTCDR task, which integrates multi-domain recommendation knowledge through domain-shared discrete semantic identifiers and unified sequence-to-sequence model.

\vspace{4pt}
\noindent\textbf{Generative Recommendation. }
With the recent advancements in generative retrieval~\cite{jin2024language,sun2024learning,zeng2024scalable,li2024learning}, there has been a growing interest in developing recommender systems based on the generative paradigm~\cite{wang2023generative,ji2024genrec,li2023gpt4rec,zhai2024actions}. Unlike traditional paradigms that typically learn a unique embedding and compute a matching score for each candidate item~\cite{wu2022survey}, generative recommendation aims to directly generate target items, thereby simplifying the recommendation process from multi-stage filtering to single-stage generation~\cite{li2024large,li2024survey,deldjoo2024review}. A crucial step in building generative recommender system is obtaining effective item identifiers. Some studies utilize the dataset's collaborative information to construct numeric identifiers for items. For instance, P5-CID~\cite{hua2023index} constructs the item co-occurrence matrix and applies matrix factorization, while SEATER~\cite{si2023generative} applies hierarchical clustering on item embeddings from traditional recommendation models. To address the lack of semantics in numeric identifiers, some approaches leverage textual metadata, such as item titles and descriptions, to represent each item~\cite{cui2022m6,yue2023llamarec}. However, due to the variability and redundancy of natural language, directly using raw text as identifiers may hinder the effectiveness of item generation. To address this issue, some pioneering studies adopt vector quantization techniques~\cite{liu2024vector,van2017neural,zeghidour2021soundstream} to tokenize an item's language representation into discrete code sequences. For instance, TIGER~\cite{rajput2023recommender} and LC-Rec~\cite{zheng2024adapting} employ RQ-VAE~\cite{lee2022autoregressive} to assign hierarchical semantic identifiers using multi-level codebooks. LETTER~\cite{wang2024learnable} and EAGER~\cite{wang2024eager} further integrate collaborative information into the semantic tokenization process. Beyond item identifier learning, some works also investigate other aspects, such as improving the training and inference efficiency~\cite{lin2024data,lin2024efficient} of generative models and utilizing generative AI tools to create and recommend entirely new items~\cite{wei2024towards,xu2024diffusion}. In this work, we focus on an under-explored research direction: solving the MTCDR task in the generative recommendan paradigm. Specifically, we thoroughly examine how to construct item identifiers and train generative recommenders to effectively incorporate multi-domain knowledge.

\begin{figure*}[t]
\centering
\includegraphics[width=\textwidth]{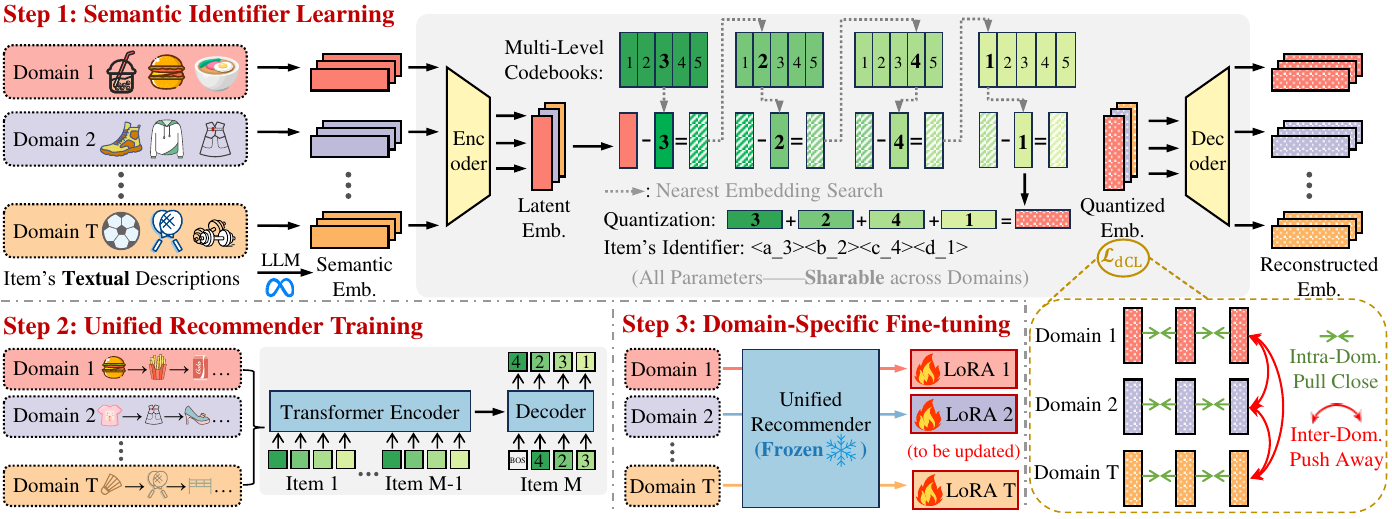}
\vspace{-12pt}
\caption{The overall framework of the proposed GMC method.}
\label{fig:framework}
\vspace{-5pt}
\end{figure*}

\section{Methodology}

In this section, we present the proposed \textbf{G}enerative recommendation framework for the \textbf{M}ulti-target \textbf{C}ross-domain recommendation task, named as \textbf{GMC}.

\subsection{Task formulation} 
Without loss of generality, we consider the sequential recommendation problem in the MTCDR task. Suppose there are $T$ parallel domains $\{\mathcal{D}_1,\cdots,\mathcal{D}_T\}$ in the considered scenario, where each domain $\mathcal{D}_t = (\mathcal{U}_t, \mathcal{I}_t, \mathcal{H}_t)$ consists of a user set $\mathcal{U}_t$, an item set $\mathcal{I}_t$, and a behavior sequence set $\mathcal{H}_t$. Specifically, each behavior sequence is formulated as the user's historically interacted items $\{i_1,i_2,\cdots,i_M\}$ (sorted chronologically), where each item $i$ is associated with a unique item ID and textual data (\eg item title, categories and brand). \textbf{The goal of MTCDR} is to improve recommendation performance of all domains simultaneously by leveraging the observed information from each domain~\cite{zhu2021cross}. To achieve this goal, the key is effectively fusing and transferring knowledge across the domains. For example, several existing studies use graph-based methods~\cite{cui2020herograph,guo2023disentangled,xu2023neural} to learn domain-shared and domain-specific knowledge. However, these methods often rely heavily on common inter-domain features, such as overlapping users (\ie $\mathcal{U}_{t_1}\cap\mathcal{U}_{t_2}\neq\varnothing$), which may be unavailable in some scenarios (\eg there may be no intersection of users in different channels of a video platform). In line with previous studies~\cite{hou2022towards,hou2023learning,li2023text}, we do not explicitly maintain user IDs of each sequence; instead, we focus on utilizing item textual data to enable information fusion and transfer between domains.

\subsection{Overview of the Approach}
Inspired by recent advances in generative recommendation, we consider solving the MTCDR task in a generative manner. To achieve this, we identify two key aspects for building generative recommender systems that enable effective information integration across multiple domains:
\begin{itemize}[leftmargin=*]
\item For \textit{item identifier learning}, we represent each item using domain-shared discrete semantic IDs, which are obtained through residual vector quantization based on item text embeddings encoded by LLMs. This provides the foundation for integrating multi-domain recommendation knowledge in model training stage.
\item For \textit{generative model training}, we model sequential recommendation as the semantic token generation task by replacing items with their corresponding identifiers. Here we leverage data from all domains to train a unified generative recommender capable of modeling recommendation knowledge across multiple domains.
\end{itemize}
Furthermore, considering the differences in data characteristics across domains, we bring closer the quantized embeddings of intra-domain items during item identifier learning, and introduce domain-specific LoRA modules into the unified recommendation model.

The overall framework of the proposed GMC approach is shown in Figure~\ref{fig:framework}, which consists of three steps (\ie semantic identifier learning, unified recommender training, and domain-specific fine-tuning). Next, we will present the details of our method.

\subsection{Semantic Identifier Learning}

A preliminary step in developing generative recommender system is assigning an identifier to each item. A straightforward approach is to represent each item by its original numeric ID~\cite{kang2018self,hua2023index}. However, this often lead to a large item vocabulary and insufficient interactions to train each embedding. In addition, the multi-domain knowledge cannot be transferred, as there are no overlapping items across domains. To properly represent each item and facilitate knowledge transfer across domains, we follow previous work~\cite{rajput2023recommender, zheng2024adapting, singh2024better} to learn semantic item identifiers. Specifically, each item is represented by a sequence of discrete semantic codewords, with each codeword shared by multiple items from different domains. The details are introduced as follows.

\textbf{Semantic Quantization. }
We first utilize large language models (LLMs) to extract the item's semantic embedding $\bm{x}\in\mathbb{R}^D$ based on its textual information, where $D$ is the dimension of LLM's token embedding. Next, we employ the RQ-VAE~\cite{lee2022autoregressive,zeghidour2021soundstream} to obtain discrete item identifiers, which recursively quantizes the residual of semantic representation and generates coarse-to-fine hierarchical indices for each item. Specifically, the semantic embedding is firstly encoded into a latent representation $\bm{z}=\text{Enc}(\bm{x}), \bm{z}\in\mathbb{R}^d$, where the encoder $\text{Enc}(\cdot)$ is implemented using a multi-layer perceptron (MLP), and $d$ denotes the reduced dimension after encoding. Then, $\bm{z}$ is processed by an $L$-level residual quantizer with codebooks $\mathcal{C}^l=\{\bm{e}_k^l\}_{k=1}^N,\ l=1,\cdots,L$, where each $\bm{e}_k^l\in\mathbb{R}^d$ is a learnable cluster center embedding, and $N$ is the codebook size. The residual quantization process is formulated as follows
\begin{align}\label{eq:rq}
    c_l &= \underset {k} { \operatorname {\arg\min} } \left\|\bm{r}_{l-1}-\bm{e}_k^l\right\|_2, \\
    \bm{r}_{l} &= \bm{r}_{l-1} - \bm{e}_{c_l}^l,
\end{align}
where we initialize with $\bm{r}_0=\bm{z}$; $\bm{r}_l$ and $c_l$ represent the residual vector and selected index during the $l$-th quantization, respectively. Intuitively, RQ-VAE searches the nearest codebook embedding $\bm{e}_{c_l}^l$ to the previous residual $\bm{r}_{l-1}$ at each level $l$, and assigns the corresponding code index $c_l$ to the item. Then the item is represented by an $L$-level quantized identifier $(c_1,\cdots,c_L)$, and the quantized latent embedding is computed as $\hat{\bm{z}} = \sum_{l=1}^{L} \bm{e}_{c_l}^l$. Afterwards, $\hat{\bm{z}}$ is passed through an MLP-based decoder to reconstruct the original semantic embedding via $\hat{\bm{x}}=\text{Dec}(\hat{\bm{z}}),\ \hat{\bm{x}}\in\mathbb{R}^D$. The overall loss for semantic quantization is computed as follows
\begin{align}
    \mathcal{L}_{\text{rqvae}} & = \mathcal{L}_{\text{recon}} + \mathcal{L}_{\text{rq}},\ \text{where}
    \label{eq:rqloss} \\
    \mathcal{L}_{\text{recon}} &= \left\|\bm{x}-\hat{\bm{x}}\right\|_2^2, \\
    \mathcal{L}_{\text{rq}} &= \sum_{l=1}^{L} \left\|\text{sg}[\bm{r}_l] - \bm{e}_{c_l}^l\right\|_2^2 + \beta \left\|\bm{r}_l - \text{sg}[\bm{e}_{c_l}^l]\right\|_2^2,
\end{align}
where $\text{sg}[\cdot]$ denotes the stop-gradient operator, and $\beta$ is a coefficient to balance the optimization strength between the reconstruction and codebook loss. Specifically, $\mathcal{L}_{\text{recon}}$ ensures the reconstruction quality of the original semantic representation, while $\mathcal{L}_{\text{rq}}$ minimizes the distance between residual vectors and codebook embeddings at each level.

\textit{Remarks.} (1) In the above semantic quantization process, we use a unified set of RQ-VAE parameters to learn identifiers for items from multiple domains. This maps item text information from different domains into a shared discrete semantic space, facilitating the efficient fusion and transfer of multi-domain recommendation knowledge by using semantic codes as a bridge. (2) Since we perform residual quantization on the latent semantic representation, the first codeword in the item identifier captures the coarsest and most critical information, while subsequent codewords work for increasingly finer granularities. This hierarchical representation aligns well with the generative nature of the recommendation model.

\textbf{Domain-aware Contrastive Loss. }
As each item $i$ is associated with its domain $t_i\in\{1,\cdots,T\}$, we aim to further leverage the domain information to improve the identifier learning. Considering that items within the same domain are generally more similar to each other than items from different domains (\eg the properties of a stationery product A are expected to be closer to another stationery product B's than a food product's), we propose to bring the quantized embeddings of intra-domain items closer together. Formally, we introduce the domain-aware contrastive loss as follows
\begin{align}\label{eq:dcl}
    \mathcal{L}_{\text{dcl}} = -\frac{1}{B} \sum_{i=1}^{B} \log \frac{\sum_{j=1}^{B} \exp(\langle\hat{\bm{z}}_i,\hat{\bm{z}}_j\rangle) \cdot \mathbb{I}[t_j = t_i]}{\sum_{j=1}^{B} \exp(\langle\hat{\bm{z}}_i,\hat{\bm{z}}_j\rangle)},
\end{align}
where $B$ denotes the batch size and $\mathbb{I}[\cdot]$ is the indicator function. Intuitively, the domain-aware contrastive loss pushes the quantized embeddings of items from the same domain within each mini-batch closer, thereby encouraging intra-domain items to exhibit similar code identifiers.

In summary, the overall training loss for semantic identifier learning is formulated as follows
\begin{align}
    \mathcal{L} = \mathcal{L}_{\text{recon}} + \mathcal{L}_{\text{rq}} + \mathcal{L}_{\text{dcl}}.
\end{align}

\subsection{Unified Recommender Training}
After semantic identifier learning, we recursively tokenize each item $i$ into its semantic identifier $\tilde{i}=(c_{i,1},\cdots,c_{i,L})$ via Eq.~\eqref{eq:rq}, and convert each user behavior log into a sequence of semantic codes. Consequently, the training set for each domain $t$ can be formulated as $\mathcal{H}_t=\{(h,y)\}$, where $h=(\widetilde{i}_1,\widetilde{i}_2,\cdots,\widetilde{i}_{M-1})$ represents the semantic code sequence of historically interacted items, and $y=\widetilde{i}_M$ denotes the next-interacted item. Following previous works~\cite{rajput2023recommender,wang2024learnable,zhu2024cost}, we model sequential recommendation as the next-token generation task, and adopt a transformer-based architecture~\cite{vaswani2017attention,raffel2020exploring} for model instantiation. To effectively model and transfer the multi-domain recommendation knowledge, we combine the training data from all domains (\ie $\mathcal{H}=\mathcal{H}_1\cup\cdots\cup\mathcal{H}_T$) to train a unified generative recommender. Formally, the model parameter $\bm{\Phi}$ is optimized as follows
\begin{align}
    \underset{\bm{\Phi}}{\max}\sum_{(h,y)\in\mathcal{H}} \sum_{l=1}^{L}\log P_{\bm{\Phi}}(y_l|h,y_{<l}).
    \label{eq:llmloss}
\end{align}

\subsection{Domain-Specific Fine-tuning}
Considering that we have combined multi-domain data to train a unified recommender, directly using the model for recommendation may overlook domain differences in data characteristics, which can be addressed through domain-specific adjustments. A straightforward approach is further training the unified model using the data $\mathcal{H}_t$ from each domain $t$. However, this may undermine the multi-domain recommendation knowledge learned by the unified model and cause the model parameters to overly focus on domain-specific information, leading to suboptimal performance. Additionally, since each domain would retain its own set of trained model parameters, the total parameter size would increase exponentially, which is both computationally intensive and time-consuming. 

As such, we propose to adopt the \textit{lightweight tuning strategy}, which updates only a small number of parameters in the unified model for each domain. This can efficiently incorporate supplementary domain-specific information (through lightweight updatable parameters) while largely preserving multi-domain knowledge (as most of the parameters are frozen). Specifically, we introduce domain-specific LoRA~\cite{hu2022lora} modules into the unified model, which involves freezing the backbone model's parameter $\bm{\Phi}$ and adding trainable rank decomposition matrices to the attention modules of the transformer layers. Formally, the LoRA parameter $\bm{\Theta}_t$ for domain $t$ is optimized as follows
\begin{align}
    \underset{\bm{\Theta}_t}{\max}\sum_{(h,y)\in\mathcal{H}_t} \sum_{l=1}^{L}\log P_{\bm{\Phi}+\bm{\Theta}_t}(y_l|h,y_{<l}),\quad t=1,\cdots,T.
    \label{eq:loraloss}
\end{align}

\textit{Inference.} After training the unified model and performing domain-specific fine-tuning, the generative recommender autoregressively generates the semantic code sequence to recommend the next item in each domain $t$ as follows
\begin{align}\label{eq:generate}
    \hat{y}_l=\underset{k\in\{1,\cdots,K\}}{\operatorname{\arg\max}} P_{\bm{\Phi}+\bm{\Theta}_t}(k|x,\hat{y}_{<l}).
\end{align}
To ensure that the generated identifier corresponds to an item in the item corpus $\mathcal{I}_t$, we follow previous works~\cite{hua2023index,xu2024openp5} to apply constrained generation on the prefix tree~\cite{de1959file} constructed from the code identifiers of items in $\mathcal{I}_t$. 

\begin{table}[t]
\caption{Comparison of our method with several related studies. "Multi-DM" denotes that the method can model multi-domain knowledge. For methods whose "Multi-DM" is \textcolor{green}{\CheckmarkBold}, "Non-OL" and "Non-AD" denote that it doesn't require overlapped users/items and auxiliary pre-training data, respectively. "GenRec" denotes Generative Recommendation.}
\vspace{-5pt}
\label{tab:comparison}
\renewcommand\arraystretch{1}
\setlength{\tabcolsep}{1mm}{
\resizebox{1\columnwidth}{!}{
\begin{tabular}{cccccc}
\toprule
\textbf{Methods} & \textbf{Text-based} & \textbf{Multi-DM} & \textbf{Non-OL} & \textbf{Non-AD} & \textbf{GenRec} \\
\midrule
\textbf{SASRec}~\cite{kang2018self} & \color{red}{\XSolidBrush} & \color{red}{\XSolidBrush} & \textbf{---} & \textbf{---} & \color{red}{\XSolidBrush} \\
\textbf{NMCDR}~\cite{xu2023neural} & \color{red}{\XSolidBrush} & \color{green}{\CheckmarkBold} & \color{red}{\XSolidBrush} & \color{green}{\CheckmarkBold} & \color{red}{\XSolidBrush} \\
\textbf{VQ-Rec}~\cite{hou2023learning} & \color{green}{\CheckmarkBold} & \color{green}{\CheckmarkBold} & \color{green}{\CheckmarkBold} & \color{red}{\XSolidBrush} & \color{red}{\XSolidBrush} \\
\textbf{P5-CID}~\cite{hua2023index} & \color{red}{\XSolidBrush} & \color{red}{\XSolidBrush} & \textbf{---} & \textbf{---} & \color{green}{\CheckmarkBold} \\
\textbf{TIGER}~\cite{rajput2023recommender} & \color{green}{\CheckmarkBold} & \color{red}{\XSolidBrush} & \textbf{---} & \textbf{---} & \color{green}{\CheckmarkBold} \\
\textbf{GMC} & \color{green}{\CheckmarkBold} & \color{green}{\CheckmarkBold} & \color{green}{\CheckmarkBold} & \color{green}{\CheckmarkBold} & \color{green}{\CheckmarkBold} \\
\bottomrule
\end{tabular}}}
\end{table}

\section{Discussion}
In this part, we briefly compare GMC with several representative methods that can be applied to address the MTCDR task, highlighting the advantages of our proposed approach.

\vspace{3.5pt}
$\bullet$ \textbf{General Recommendation Approaches} such as GRU4Rec~\cite{hidasi2015session}, SASRec~\cite{kang2018self} and HGN~\cite{ma2019hierarchical} utilize raw item IDs to construct and train the recommender. Since item IDs in different domains are not sharable, separate models must be trained for each domain, preventing the integration of multi-domain knowledge.

$\bullet$ \textbf{Traditional MTCDR Approaches} such as HeroGRAPH~\cite{cui2020herograph}, NMCDR~\cite{xu2023neural} and DR-MTCDR~\cite{guo2023disentangled} rely on overlapping users, items, or tags across domains to fuse and transfer multi-domain recommendation knowledge. These methods, however, cannot capture cross-domain information in more general settings where such overlapping information is unavailable.

$\bullet$ \textbf{Context-Aware Cross-Domain Approaches} such as UniSRec~\cite{hou2022towards}, VQ-Rec~\cite{hou2023learning} and RecFormer~\cite{li2023text} utilize textual information to model user behaviors and items as universal semantic representations. To learn high-quality representations, they rely on additional source domain data for model pre-training, which is of large size and often unavailable. Without such pre-training data, the recommendation performance would be sub-optimal.

$\bullet$ \textbf{Generative Approaches} such as P5-CID~\cite{hua2023index}, TIGER~\cite{rajput2023recommender} and LC-Rec~\cite{zheng2024adapting} represent each item with an identifier and generate the item identifier tokens autoregressively. Although generative recommendation has emerged as the next-generation paradigm, prior works primarily focus on single-domain settings. The application of generative approaches to the MTCDR task remains under-explored.

\vspace{3.5pt}
In contrast, our proposed GMC method maps the textual information of items from multiple domains into a shared discrete semantic space using codewords, and then trains a unified generative recommender to model multi-domain recommendation knowledge. By aligning multi-domain item semantics into a unified codebook-driven latent space, GMC facilitates structured knowledge transfer through codeword-guided generation, while the unified architecture inherently mitigates information fragmentation typical of domain-isolated recommenders. This enables GMC to harness cross-domain information more effectively than traditional non-generative approaches, achieving more promising performance.

The comparison of these approaches is summarized in Table~\ref{tab:comparison}.

\section{Experiments}

In this section, we first set up the experiments, and then present the overall performance comparison and in-depth analysis.

\begin{table}[t]
\centering
\caption{Statistics of the processed datasets. ``\textbf{Avg.}\emph{len}'' denotes the average length of item sequences.}
\vspace{-8pt}
\label{tab:statistics}
\renewcommand\arraystretch{1}
\setlength{\tabcolsep}{1mm}{
\resizebox{1\columnwidth}{!}{
\begin{tabular}{lccccc}
\toprule
\textbf{Datasets} & \textbf{\#User} & \textbf{\#Item} & \textbf{\#Interaction} & \textbf{Sparsity} & \textbf{Avg.}\textit{len} \\
\midrule
\textbf{Scientific} & 8,317 & 4,344 & 58,492 & 99.84\% & 7.03 \\
\textbf{Pantry} & 13,073 & 4,897 & 126,693 & 99.80\% & 9.69 \\
\textbf{Instruments} & 24,772 & 9,922 & 206,153 & 99.92\% & 8.32 \\
\textbf{Arts} & 45,141 & 20,956 & 390,832 & 99.96\% & 8.66 \\
\textbf{Office} & 86,554 & 25,847 & 675,882 & 99.97\% & 7.81 \\
\bottomrule
\end{tabular}}}
\vspace{-5pt}
\end{table}

\subsection{Experimental Setup}

\subsubsection{Datasets} 
We conduct experiments on a MTCDR task, which consists of five subsets of Amazon Review Dataset\footnote{\url{https://cseweb.ucsd.edu/\~jmcauley/datasets/amazon\_v2/}}~\cite{ni2019justifying}, including \textit{Scientific}, \textit{Pantry}, \textit{Instruments}, \textit{Arts} and \textit{Office}. Following previous works~\cite{hou2023learning,zheng2024adapting}, we exclude items without title, and apply 5-core filtering to filter out users and items with few interactions. Then we sort each user's interacted items chronologically to construct the behavior sequences, and the maximal number of items is uniformly set to 20. For each item, we concatenate its \textit{title}, \textit{brand} and \textit{categories} fields to obtain the textual description. The statistics of the processed datasets are summarized in Table~\ref{tab:statistics}.

\subsubsection{Baselines} 
We compare the proposed approach with the following four groups of baselines, including methods using only item IDs, methods using both IDs and text features, methods based on model pre-training, and generative recommendation methods.

\begin{enumerate}[label=(\arabic*),leftmargin=*]
\item \textit{ID-only methods:}
\begin{itemize}
 \item {\textbf{SASRec}}~\cite{kang2018self} is a self-attention based sequential model which captures long-term user preferences.
 \item {\textbf{BERT4Rec}}~\cite{sun2019bert4rec} employs the deep bidirectional self-attention to model user behavior sequences with cloze objectives.
 \item {\textbf{HGN}}~\cite{ma2019hierarchical} applies hierarchical gating networks to model both the long-term and short-term user interests.
 \item {\textbf{GRU4Rec}}~\cite{hidasi2015session} utilizes stacked gated recurrent units to capture sequential dependencies within user interactions.
\end{itemize}

\item \textit{ID-text methods:}
\begin{itemize}
 \item {\textbf{FDSA}}~\cite{zhang2019feature} adopts feature-level self-attention blocks to leverage the item's attribute information. 
 \item {\textbf{S$^3$-Rec}}~\cite{zhou2020s3} leverages the mutual information maximization principle to pre-train the sequential recommendation model with four self-supervised objectives.
\end{itemize}

\item \textit{Pre-training-based methods:}
\begin{itemize}
 \item {\textbf{UniSRec}}~\cite{hou2022towards} utilizes item texts to learn universal item and sequence representations on pre-training data, and adapts an MoE-enhanced adaptor for new domain adaptation.
 \item {\textbf{VQ-Rec}}~\cite{hou2023learning} employs a "text$\Rightarrow$code$\Rightarrow$representation" scheme to learn vector-quantized item representations for transferable sequential recommendation.
 \item {\textbf{RecFormer}}~\cite{li2023text} transforms items and interactions into text sequences and then leverages the behavior-tuned model to obtain their representations.
\end{itemize}
To align these baselines with the MTCDR setting, instead of pre-training the model with auxiliary source-domain datasets as described in their original papers, we use a combination of our five selected target datasets for pre-training.

\item \textit{Generative methods:}
\begin{itemize}
 \item {\textbf{P5-CID}}~\cite{geng2022recommendation,hua2023index} injects collaborative signals into item identifiers by spectral clustering on item's co-occurrence graph.
 \item {\textbf{TIGER}}~\cite{rajput2023recommender} first quantizes item's semantic information into codebook identifiers, and then adopts the generative retrieval paradigm for recommendation.
 \item {\textbf{IDGenRec}}~\cite{tan2024idgenrec} represents each item as a concise textual ID by training an ID generator alongside the recommender.
\end{itemize}
\end{enumerate}

\begin{table*}[t]
\centering
\caption{The performance of baselines and our proposed method. The optimal, sub-optimal and third-best results are highlighted in \textbf{bold}, \underline{underline} and \uwave{wave}, respectively. "Imp." denotes the relative improvement of GMC compared to the strongest baseline. The value of AVERAGE is calculated as a user-weighted average of the metrics from the five domains.} 
\vspace{-5pt}
\label{tab:overall}
\renewcommand\arraystretch{1.28}
\setlength{\tabcolsep}{0.6mm}{
\resizebox{\textwidth}{!}{
\begin{tabular}{l|l|cccc|cc|ccc|ccc|cc}
\toprule
\textbf{Dataset} & \textbf{Metric} & \textbf{SASRec} & \textbf{BERT4Rec} & \textbf{HGN} & \textbf{GRU4Rec} & \textbf{FDSA} & \textbf{S$^3$-Rec} & \textbf{UniSRec} & \textbf{VQ-Rec} & \textbf{RecFormer} & \textbf{P5-CID} & \textbf{TIGER} & \textbf{IDGenRec} & \textbf{GMC} & \textbf{Imp.} \\ 
\midrule
 & \textbf{R@5} & 0.0753 & 0.0528 & 0.0616 & 0.0493 & 0.0673 & 0.0641 & {\ul 0.0836} & \textbf{0.0839} & 0.0760 & 0.0339 & 0.0632 & 0.0635 & \uwave{0.0791} & -- \\ 
 & \textbf{R@10} & 0.1017 & 0.0697 & 0.0916 & 0.0690 & 0.0860 & 0.0866 & {\ul 0.1159} & \textbf{0.1202} & 0.1009 & 0.0537 & 0.0871 & 0.0850 & \uwave{0.1091} & -- \\ 
 & \textbf{N@5} & 0.0488 & 0.0391 & 0.0394 & 0.0364 & {\ul 0.0528} & 0.0429 & \uwave{0.0519} & 0.0498 & \uwave{0.0519} & 0.0221 & 0.0457 & 0.0478 & \textbf{0.0556} & 5.34\% \\
\multirow{-4}{*}{\textbf{Scientific}} & \textbf{N@10} & 0.0573 & 0.0445 & 0.0491 & 0.0427 & 0.0588 & 0.0502 & {\ul 0.0623} & \uwave{0.0616} & 0.0598 & 0.0285 & 0.0534 & 0.0547 & \textbf{0.0652} & 5.79\% \\
\midrule
 & \textbf{R@5} & 0.0315 & 0.0216 & 0.0262 & 0.0261 & 0.0261 & 0.0223 & \uwave{0.0375} & {\ul 0.0402} & 0.0306 & 0.0187 & 0.0316 & 0.0313 & \textbf{0.0422} & 5.04\% \\
 & \textbf{R@10} & 0.0539 & 0.0367 & 0.0480 & 0.0443 & 0.0407 & 0.0422 & \uwave{0.0643} & \textbf{0.0667} & 0.0531 & 0.0317 & 0.0516 & 0.0481 & {\ul 0.0656} & -- \\ 
 & \textbf{N@5} & 0.0165 & 0.0132 & 0.0165 & 0.0163 & 0.0173 & 0.0121 & {\ul 0.0218} & \uwave{0.0209} & 0.0183 & 0.0122 & 0.0199 & 0.0195 & \textbf{0.0262} & 25.41\% \\
\multirow{-4}{*}{\textbf{Pantry}} & \textbf{N@10} & 0.0237 & 0.0180 & 0.0235 & 0.0221 & 0.0220 & 0.0185 & {\ul 0.0305} & \uwave{0.0294} & 0.0255 & 0.0164 & 0.0263 & 0.0249 & \textbf{0.0338} & 14.89\% \\
\midrule
 & \textbf{R@5} & 0.0817 & 0.0791 & 0.0826 & 0.0863 & 0.0861 & 0.0730 & \uwave{0.0898} & {\ul 0.0912} & 0.0778 & 0.0775 & 0.0876 & 0.0857 & \textbf{0.0980} & 7.47\% \\
 & \textbf{R@10} & 0.1087 & 0.0970 & 0.1054 & 0.1067 & 0.1061 & 0.0953 & \uwave{0.1201} & {\ul 0.1209} & 0.0966 & 0.0974 & 0.1089 & 0.1034 & \textbf{0.1224} & 1.24\% \\
 & \textbf{N@5} & 0.0549 & 0.0670 & 0.0680 & 0.0727 & 0.0735 & 0.0545 & 0.0662 & 0.0637 & 0.0542 & 0.0665 & \uwave{0.0743} & {\ul 0.0745} & \textbf{0.0828} & 11.08\% \\
\multirow{-4}{*}{\textbf{Instruments}} & \textbf{N@10} & 0.0636 & 0.0727 & 0.0753 & 0.0793 & 0.0799 & 0.0617 & 0.0759 & 0.0733 & 0.0603 & 0.0729 & {\ul 0.0812} & \uwave{0.0802} & \textbf{0.0906} & 11.58\% \\
\midrule
 & \textbf{R@5} & 0.0751 & 0.0605 & 0.0621 & 0.0743 & 0.0748 & 0.0646 & \uwave{0.0803} & {\ul 0.0823} & 0.0746 & 0.0621 & 0.0782 & 0.0773 & \textbf{0.0860} & 4.55\% \\
 & \textbf{R@10} & 0.1007 & 0.0780 & 0.0865 & 0.0959 & 0.0953 & 0.0883 & {\ul 0.1126} & \textbf{0.1155} & 0.0992 & 0.0803 & 0.1031 & 0.0970 & \uwave{0.1103} & -- \\ 
 & \textbf{N@5} & 0.0522 & 0.0486 & 0.0472 & 0.0598 & 0.0599 & 0.0448 & 0.0555 & 0.0558 & 0.0521 & 0.0506 & \uwave{0.0629} & {\ul 0.0638} & \textbf{0.0690} & 8.09\% \\
\multirow{-4}{*}{\textbf{Arts}} & \textbf{N@10} & 0.0604 & 0.0543 & 0.0551 & 0.0668 & 0.0665 & 0.0524 & 0.0658 & 0.0665 & 0.0601 & 0.0564 & {\ul 0.0709} & \uwave{0.0701} & \textbf{0.0768} & 8.34\% \\
\midrule
 & \textbf{R@5} & 0.0915 & 0.0846 & 0.0803 & 0.0946 & 0.0943 & 0.0871 & 0.0961 & 0.0949 & 0.0896 & 0.0916 & \uwave{0.1000} & {\ul 0.1043} & \textbf{0.1063} & 1.91\% \\
 & \textbf{R@10} & 0.1098 & 0.0960 & 0.0972 & 0.1112 & 0.1087 & 0.1056 & \uwave{0.1192} & 0.1174 & 0.1074 & 0.1041 & 0.1180 & {\ul 0.1198} & \textbf{0.1252} & 4.58\% \\
 & \textbf{N@5} & 0.0708 & 0.0745 & 0.0658 & 0.0817 & 0.0810 & 0.0682 & 0.0722 & 0.0710 & 0.0666 & 0.0814 & \uwave{0.0857} & {\ul 0.0902} & \textbf{0.0912} & 1.11\% \\
\multirow{-4}{*}{\textbf{Office}} & \textbf{N@10} & 0.0767 & 0.0781 & 0.0713 & 0.0870 & 0.0856 & 0.0742 & 0.0797 & 0.0782 & 0.0724 & 0.0854 & \uwave{0.0914} & {\ul 0.0952} & \textbf{0.0973} & 2.22\% \\
\midrule
 & \textbf{R@5} & 0.0808 & 0.0716 & 0.0712 & 0.0811 & 0.0819 & 0.0736 & 0.0863 & \uwave{0.0867} & 0.0792 & 0.0741 & 0.0860 & {\ul 0.0876} & \textbf{0.0940} & 7.34\% \\
 & \textbf{R@10} & 0.1028 & 0.0860 & 0.0917 & 0.0998 & 0.0989 & 0.0942 & \uwave{0.1135} & {\ul 0.1138} & 0.0995 & 0.0894 & 0.1066 & 0.1048 & \textbf{0.1159} & 1.85\% \\
 & \textbf{N@5} & 0.0588 & 0.0607 & 0.0565 & 0.0680 & 0.0686 & 0.0550 & 0.0625 & 0.0615 & 0.0570 & 0.0636 & \uwave{0.0716} & {\ul 0.0742} & \textbf{0.0780} & 5.14\% \\
\multirow{-4}{*}{\textbf{AVERAGE}} & \textbf{N@10} & 0.0659 & 0.0653 & 0.0632 & 0.0740 & 0.0740 & 0.0617 & 0.0712 & 0.0702 & 0.0635 & 0.0686 & \uwave{0.0782} & {\ul 0.0797} & \textbf{0.0850} & 6.67\% \\
\bottomrule
\end{tabular}}}
\end{table*}

\subsubsection{Evaluation Settings} 
Following previous works~\cite{kang2018self,hou2022towards,tan2024idgenrec}, we adopt the leave-one-out strategy to split datasets: for each user's interaction records, the last item is used as the test data, the second latest item is used for validation, and the remaining item sequence are used for training. To evaluate the recommendation performance, we rank the target item over the entire item set~\cite{krichene2020sampled}, and adopt two widely-used metrics, Recall@$K$ and NDCG@$K$~\cite{jarvelin2002cumulated}, with $K\in\{5,10\}$. We report the average metric scores of all test users in each dataset, and the average performance of all datasets.

\subsubsection{Implementation Details} 
To obtain item identifiers, we first adopt LLaMA-3.1 model\footnote{\url{https://huggingface.co/meta-llama/Llama-3.1-8B-Instruct}}~\cite{touvron2023llama,dubey2024llama} to encode each item's textual description into a latent vector with dimension $D=4096$, and then perform semantic tokenization through an RQ-VAE model, whose encoder and decoder are MLPs with ReLU activations. Following previous works~\cite{rajput2023recommender,zheng2024adapting,wang2024learnable}, the level of codebook is set to $L=4$, and each codebook consists of $N=256$ learnable embeddings with dimension $d=32$. When computing the total loss, we use $\beta=0.25$. The model is trained for 10k epochs by the AdamW optimizer~\cite{loshchilov2017decoupled}, with a learning rate of 0.001 and batch size of 1024.

For the generative recommender training, we adopt the open-sourced T5 transformer\footnote{\url{https://huggingface.co/google-t5/t5-small}}~\cite{raffel2020exploring} as the backbone sequence-to-sequence model. In the unified recommender training stage, we train the model for 200 epochs with the learning rate set to 2e-3 and batch size set to to 1024. In the domain-specific fine-tuning stage, we set the learning rate to 5e-4 and batch size to 512, and adopt the early-stop strategy to prevent overfitting. In the item generation phase, the beam size is uniformly set to 20.

\subsection{Overall Performance}

We compare the proposed GMC with baseline methods across five benchmark datasets, treating each as a distinct domain, and present the overall performance in Table~\ref{tab:overall}. From the results, we draw the following observations:

\begin{itemize}[leftmargin=*]
\item Among the traditional ID-based single-domain methods, the text-enhanced approach FDSA outperforms methods that rely solely on explicit IDs (\ie SASRec, BERT4Rec, HGN and GRU4Rec) on several datasets. This indicates that integrating textual information of items into ID-based models can effectively improve recommendation performance.
\item For recent generative recommendation approaches, TIGER and IDGenRec consistently achieves better performance than P5-CID. This can be attributed to the way P5-CID constructs collaborative item identifiers based on the item co-occurrence graph, which inadequately captures content information. In contrast, TIGER and IDGenRec build identifiers leveraging semantic information, which is more benericial for generative recommendation.
\item Compared to other baselines, pre-training-based methods UniSRec and VQ-Rec achieves superior recommendation performance, especially on small-scale datasets (\ie Pantry and Scientific). This demonstrates the effectiveness of leveraging multi-domain recommendation data to enhance single-domain recommendation performance through pre-training.
\item Our proposed method GMC achieves the best performance in almost all cases, especially in terms of the ranking-ability-related NDCG metrics. This superior performance can be attributed to GMC's ability to effectively integrate and transfer multi-domain knowledge through domain-shared discrete semantic codes, and the generative paradigm also empowers GMC the strong recommendation ability.
\end{itemize}

\subsection{Ablation Study}

\begin{table*}
\centering
\caption{Ablation study of GMC. "RI" indicates the relative improvement of GMC over its variants.}
\vspace{-6pt}
\label{tab:ablation}
\renewcommand\arraystretch{1.3}
\setlength{\tabcolsep}{0.7mm}{
\resizebox{1.0\textwidth}{!}{
\begin{tabular}{@{}lcccccccccccccccc@{}}
\toprule
\multirow{2}{*}{\textbf{Variants}} & \multicolumn{4}{c}{\textbf{Scientific}} & \multicolumn{4}{c}{\textbf{Pantry}} & \multicolumn{4}{c}{\textbf{Instruments}} & \multicolumn{4}{c}{\textbf{Office}} \\
\cmidrule(lr){2-5} \cmidrule(lr){6-9} \cmidrule(lr){10-13} \cmidrule(lr){14-17}
 & R@5 & RI & \ \ N@5 & RI & R@5 & RI & \ \ N@5 & RI & R@5 & RI & \ \ N@5 & RI & R@5 & RI & \ \ N@5 & RI \\
\midrule
(0) GMC & \ \ \textbf{0.0791} & - & \ \ {\ul 0.0556} & - & \ \ \textbf{0.0422} & - & \ \ \textbf{0.0262} & - & \ \ \textbf{0.0980} & - & \ \ {\ul 0.0828} & - & \ \ {\ul 0.1063} & - & \ \ {\ul 0.0912} & - \\
\cmidrule(r){1-1}
(1) \textit{w/o} sharable codebook & \ \ 0.0632 & 25.1\% & \ \ 0.0457 & 21.7\% & \ \ 0.0316 & 33.7\% & \ \ 0.0199 & 31.8\% & \ \ 0.0876 & 11.9\% & \ \ 0.0743 & 11.4\% & \ \ 0.1000 & 6.3\% & \ \ 0.0857 & 6.5\% \\
(2) \textit{w/o} unified recommender & \ \ 0.0608 & 30.0\% & \ \ 0.0441 & 26.1\% & \ \ 0.0283 & 49.2\% & \ \ 0.0177 & 48.0\% & \ \ 0.0878 & 11.6\% & \ \ 0.0752 & 10.0\% & \ \ 0.1032 & 3.0\% & \ \ 0.0884 & 3.2\% \\
\cmidrule(r){1-1}
(3) \textit{w/o} $\mathcal{L}_{\text{dcl}}$ & \ \ 0.0754 & 4.9\% & \ \ 0.0547 & 1.7\% & \ \ 0.0388 & 8.9\% & \ \ 0.0246 & 6.6\% & \ \ 0.0942 & 4.1\% & \ \ 0.0806 & 2.7\% & \ \ 0.1061 & 0.2\% & \ \ 0.0908 & 0.4\% \\
(4) \textit{w/o} lora finetuning & \ \ 0.0765 & 3.5\% & \ \ 0.0537 & 3.6\% & \ \ {\ul 0.0401} & 5.3\% & \ \ {\ul 0.0248} & 5.8\% & \ \ {\ul 0.0974} & 0.6\% & \ \ 0.0827 & 0.1\% & \ \ 0.1062 & 0.1\% & \ \ 0.0910 & 0.2\% \\
(5) finetune all parameters & \ \ {\ul 0.0788} & 0.4\% & \ \ \textbf{0.0560} & - & \ \ 0.0395 & 7.0\% & \ \ 0.0245 & 6.8\% & \ \ 0.0972 & 0.8\% & \ \ \textbf{0.0829} & - & \ \ \textbf{0.1074} & - & \ \ \textbf{0.0923} & - \\
\bottomrule
\end{tabular}}}
\vspace{0pt}
\end{table*}

\begin{figure*}[t]

\centering
\includegraphics[width=0.83\textwidth]{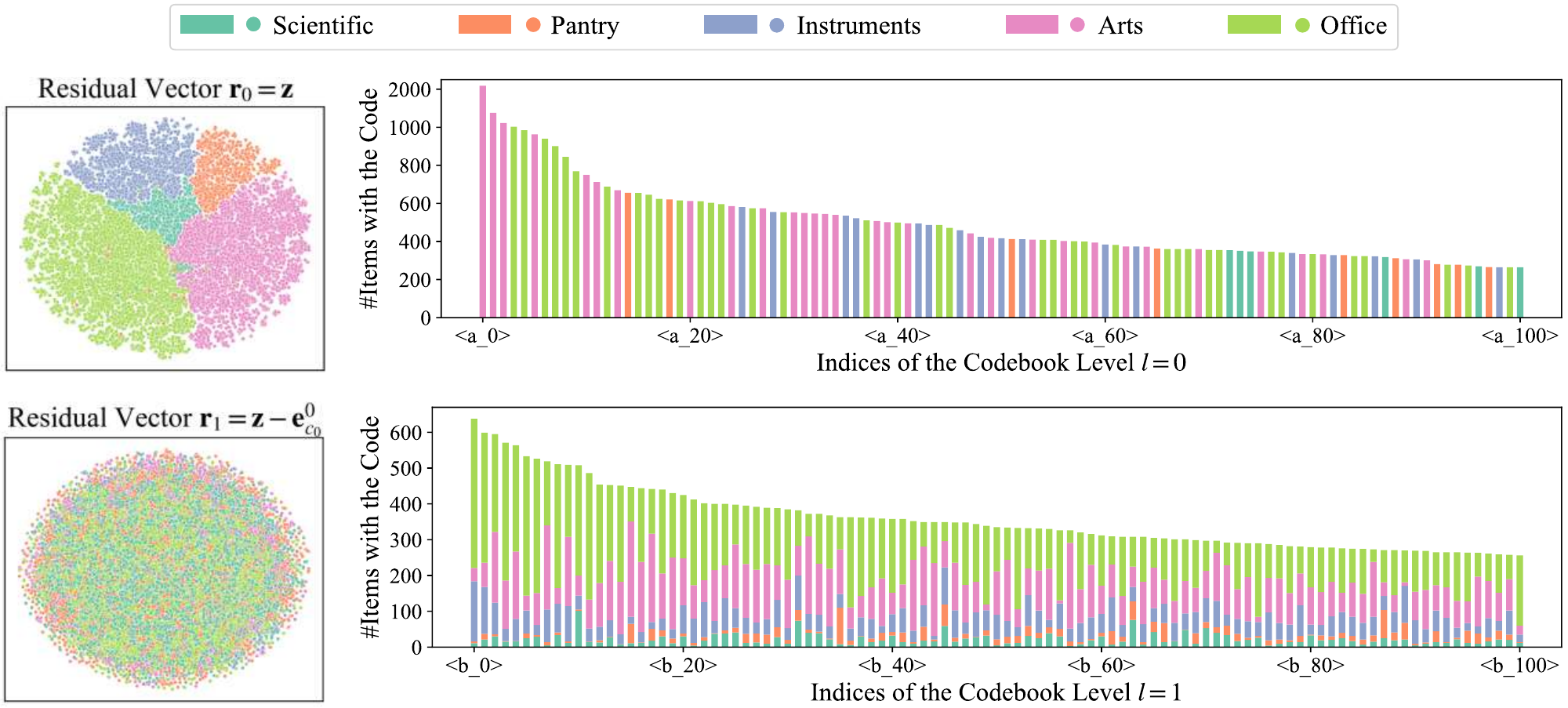}
\caption{Distribution of residual vectors and semantic codes in item identifier learning. Each bar consists of items with the corresponding semantic code, where each code can be shared by items from different domains. We visualize the 100 most frequent codes, and the distribution trends of remaining codes are similar.}
\label{fig:code_dist}
\end{figure*}

In this section, we analyze the impact of each specific design in our proposed method on recommendation performance. We report the performance comparison between the default GMC and its five variants in Table~\ref{tab:ablation}.

(1) In \underline{\textit{w/o} sharable codebook}, the item identifiers for different domains are learned separately, preventing the recommenders from leveraging cross-domain knowledge. This variant results in a significant performance drop compared to GMC, particularly on small-scale datasets Scientific and Pantry, which demonstrates the importance of modeling and transferring multi-domain recommendation knowledge for enhancing performance in each domain.

(2) \underline{\textit{w/o} unified recommender} learns domain-shared item identifiers in the same way as GMC, but trains separate generative recommendation models for each domain. We observe that the performance of this variant drops sharply, and it performs even worse on Scientific and Pantry than Variant (1). This suggests that learning unified codebooks for multiple domains only realizes its full cross-domain potential when paired with a unified recommender; otherwise, the quality of domain-shared item identifiers may be inferior to those learned separately for each domain.

(3) \underline{\textit{w/o} $\mathcal{L}_{\text{dcl}}$} removes the domain-aware contrastive loss (Eq.~\eqref{eq:dcl}) in semantic identifier learning, which also degrades the recommendation performance. This phenomenon validates the effectiveness of the proposed $\mathcal{L}_{\text{dcl}}$ loss, which leverages domain information to enhance identifier learning.

(4) \underline{\textit{w/o} lora fine-tuning} directly adopts the unified generative model for recommendation. The results demonstrate that domain-specific fine-tuning can further improve the recommendation performance, particularly on small-scale domains Scientific and Pantry.

(5) \underline{fine-tune all parameters} involves updating all parameters of the unified model for each domain. This variant generally performs worse than the default lightweight tuning strategy in GMC, with a particularly sharp performance drop on the Pantry domain. This indicates that fully training the unified recommender on each domain may undermine multi-domain knowledge and lead to overfitting. In addition, the parameter size of the domain-specific LoRA module (0.17M) constitutes only 3.7\% of the entire model's (4.3M), demonstrating the parameter efficiency of the proposed strategy. We also note that this variant consistently outperforms GMC on Office, likely because its relatively large data size benefits more from full model updating compared to tuning only the LoRA modules.

In summary, as shown in Table~\ref{tab:ablation} and the above analysis, all the designs in GMC can improve the recommendation performance, especially on small-scale domains. This further validates their effectiveness in modeling and transferring cross-domain knowledge.

\subsection{In-Depth Analysis}

\subsubsection{Analysis on Code Assignment Distribution across Domains}
In this part, we investigate the item identifier sharing mechanism of GMC. As the $l$-th codeword is obtained via nearest embedding search on the $l$-th codebook using the residual vector $\bm{r}_l$, we visualize the item's residual vectors during quantization using t-SNE, and examine the assignment of codes to items across different domains. The observations and analysis\footnote{As the trend of distributions of the 3rd and 4th level is similar to the 2nd level's, we only visualize the distributions of the first two levels due to space limit.} based on Figure~\ref{fig:code_dist} are as follows.

$\bullet$ During the first-level quantization ($l=0$), the item's residual vectors $\bm{r}_0$ (\ie the latent semantic representation $\bm{z}$) are layered according to domains. Meanwhile, the semantic codes are separated, indicating that there are no overlapping codes across domains. As the domain information serves as the most coarse-grained feature of an item, and the proposed domain-aware contrastive loss further brings intra-domain item representations closer, $\bm{r}_0$ will naturally cluster items based on domain information. Building on this, the first-level codebook maps items from each domain into finer-grained subgroups, ultimately resulting in separated codewords across domains. Therefore, the first semantic codes mainly capture the domain-specific item information, which is not explicitly shared across multiple domains.

$\bullet$ During the second-level quantization ($l=1$), the residual vectors $\bm{r}_1$ from different domains are mixed, and each codeword is shared across multiple domains. Based on the previous analysis, since the residual vector $\bm{r}_1=\bm{z}-\bm{e}_{c_0}^0$ subtracts the first-level codebook embedding (which encodes domain-specific information) from the semantic representation, the resulting residual vectors $\bm{r}_1$ primarily capture domain-agnostic semantic information. Consequently, the semantic codes are shared by items from different domains, facilitating the fusion and transfer of multi-domain recommendation knowledge.

\subsubsection{Similarity Analysis \wrt Item Identifiers}

\begin{figure}[t]

\centering
\includegraphics[width=\columnwidth]{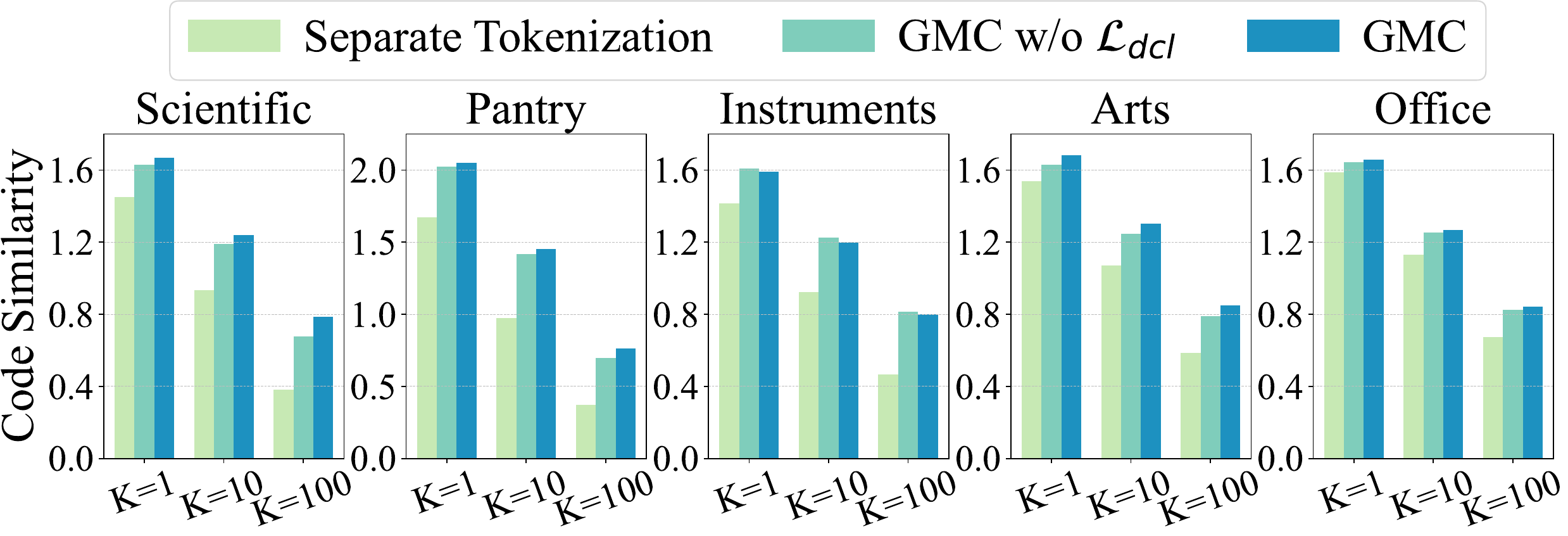}
\vspace{-15pt}
\caption{Code Similarity of Item Identifiers.}
\label{fig:code_similarity}
\vspace{-8pt}
\end{figure}

In this part, we show that the domain-shared codebook and domain-aware contrastive loss designs can benefit the semantic identifier learning from improving the quality of tokenized identifiers. Specifically, we investigate whether items with similar semantics exhibit similar code identifiers. For each item $i$ in domain $t$, we retrieve its $K$ most similar items based on the distance between their original semantic embeddings. We then compute the average overlap of codewords between these $K$ items and item $i$, where the item identifiers are learned using different strategies. Specifically, a higher overlap indicates that the learned identifiers better reflect the semantic similarity between items. The average results for all items in each domain with varying $K$ values are reported in Figure~\ref{fig:code_similarity}. We can find that compared to constructing item identifiers separately in different domains, learning domain-shared identifiers consistently improves the code similarity of semantically similar items in each domain, which is beneficial for generative recommendation. Additionally, the domain-aware contrastive loss further enhances the desired code similarity, demonstrating its effectiveness.

\subsubsection{Performance Comparison on Different Multi-Domain Scenarios}

\begin{figure}[t]

\centering
\includegraphics[width=\columnwidth]{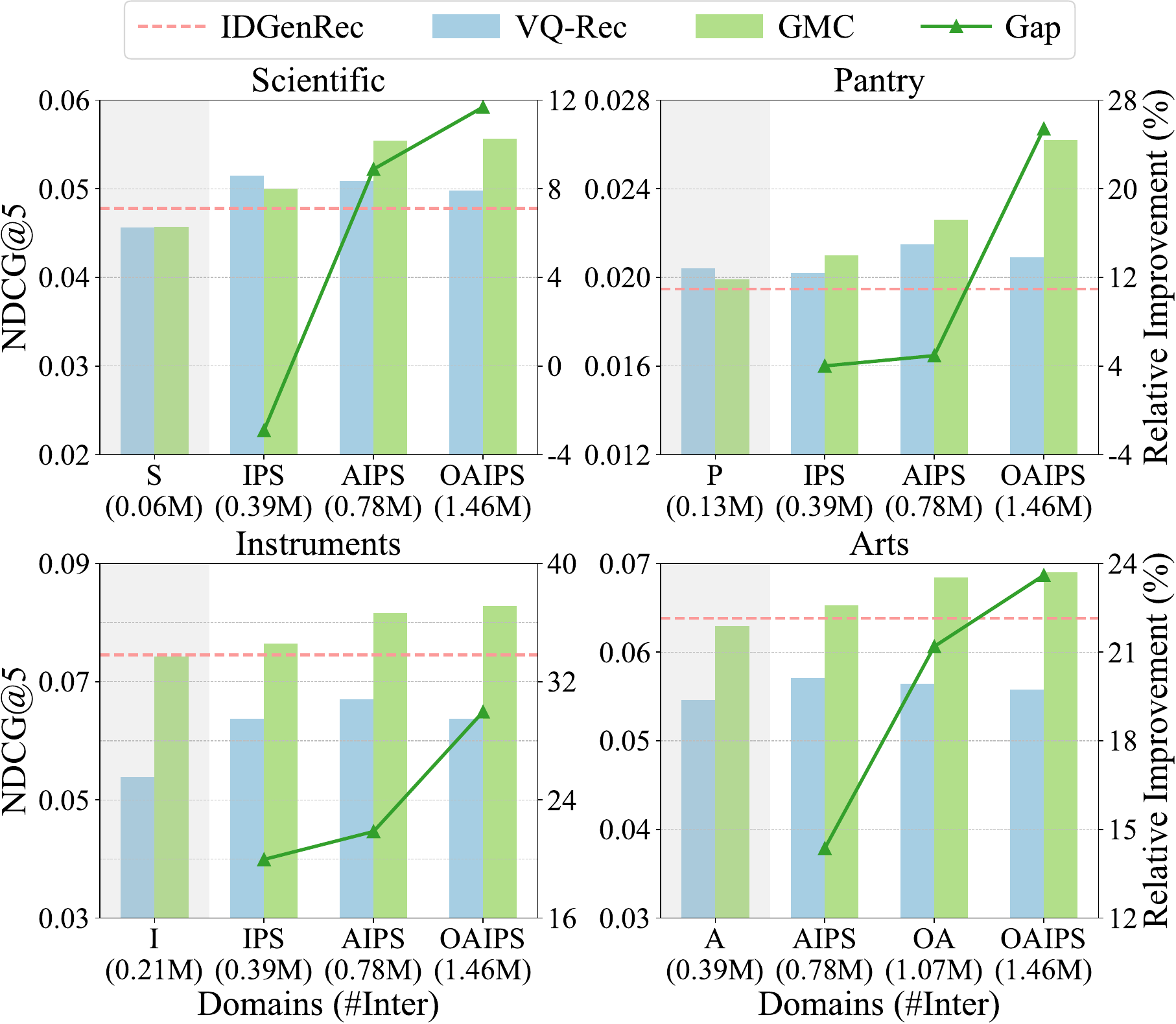}
\vspace{-12pt}
\caption{Performance comparison in different multi-domain settings. O/A/I/P/S are short for Office / Arts / Instruments / Pantry / Scientific, respectively. "Gap" denotes the relative performance improvement of GMC over VQ-Rec. 'M' indicates that the number of interactions is measured in millions. The area in shadow denotes the single-domain settings.}
\label{fig:various_domains}
\end{figure}

In this part, we further compare and analyze the effectiveness of GMC under various multi-domain settings. Specifically, we create multi-domain recommendation scenarios with different data sizes by selecting and combining datasets. We then evaluate the recommendation performance of our proposed GMC and VQ-Rec under these settings, where both methods firstly pre-train the recommender on all domains and then fine-tune it on each target domain. The experimental results are shown in Figure~\ref{fig:various_domains}, leading to the following observations. (1) Compared to training the recommender on single-domain data, multi-domain learning enhances the performance of both GMC and VQ-Rec. Moreover, GMC consistently outperforms the single-domain baseline, IDGenRec, across various multi-domain settings. These results demonstrate the effectiveness of modeling multi-domain knowledge for improving recommendation performance. (2) As the total data size across multiple domains increases, the performance of VQ-Rec does not consistently improve. In contrast, GMC shows consistent improvement, along with a growing relative performance advantage over VQ-Rec. This further validates that GMC can effectively integrate and transfer knowledge across multiple domains to enhance the recommendation performance.

\section{Conclusion}

In this paper, we propose GMC, a novel generative recommendation framework for solving the multi-target cross-domain recommendation problem. The key idea of GMC lies in leveraging semantic quantization to generate domain-shared discrete code identifiers, which serve as a bridge for integrating cross-domain knowledge by training a unified generative recommendation model. To further consider the domain information, we introduce a domain-aware contrastive loss to enhance intra-domain quantized embedding alignment in identifier learning, and apply domain-specific lightweight fine-tuning to the well-trained unified recommender. Extensive experiments on five real-world datasets validate the effectiveness of GMC, achieving superior performance compared to other competitive baselines.

In the future, we plan to explore the following research directions: (1) As incorporating collaborative information in semantic tokenization~\cite{wang2024learnable,wang2024eager} may enhance the recommendation performance, it's valuable to investigate how to transfer collaborative knowledge from different domains during the domain-unified item tokenization process. (2) Similar to most prior studies, GMC treats identifier learning and recommender training as independent tasks, and effectively integrating these two stages~\cite{jin2024language,liu2024end} in building generative MTCDR frameworks is a promising direction. (3) In scenarios where overlapping users or items exist across domains, leveraging this shared information could enhance recommendation performance, thus integrating domain-shared entities into the GMC framework is another worth-exploring topic.

\newpage

\balance
\bibliographystyle{ACM-Reference-Format}
\bibliography{sample-base}




\end{document}